\documentstyle[11pt, aaspp4]{article}

\slugcomment{Novermber 20, 1995}

\lefthead{Pak, Jaffe, \& Keller}
\righthead{\hh\ from the Inner 400 pc of the Galaxy}

\def\hii{\hbox{H\,{\scriptsize II}}}

\def\thirteenco{\hbox{$^{13}$CO}}

\def\hh{\hbox{H$_2$}}
\def\vone{\hbox{$v = 1 \rightarrow 0$ S(1)}}
\def\vtwo{\hbox{$v = 2 \rightarrow 1$ S(1)}}

\def\msun{\hbox{$M_{\odot}$}}
\def\lsun{\hbox{$L_{\odot}$}}

\def\cmv{\hbox{cm$^{-3}$}}

\def\kms{\hbox{km\,s$^{-1}$}}
\def\ergintensity{\hbox{ergs\,s$^{-1}$\,cm$^{-2}$\,sr$^{-1}$}}
\def\ergflux{\hbox{ergs\,s$^{-1}$\,cm$^{-2}$}}

\def\ak{\hbox{${\rm A_K}$}}

\def\sgra{\hbox{Sgr~A}}

\def\sgrbtwo{\hbox{Sgr~B$_2$}}

\begin{document}

%%%%%%%%%%%% Front Page %%%%%%%%%%%%
.
\vspace{2.5cm}

\begin{center}
  {\bf \Large
    H$_2$ EMISSION FROM \\
    THE INNER 400 PC OF THE GALAXY }

  \vspace{1cm}
  {\large
    Soojong Pak, D. T. Jaffe, and L. D. Keller

    Department of Astronomy, The University of Texas\\
    Austin, TX 78712
  }

  \vspace{5cm}
  To appear in {\it the Astrophysical Jounal Letters}

  vol. 457 on Jan. 20, 1996 
\end{center}
\clearpage
%%%%%%%%%%%%%%%%%%%%%%%%%%%%%%%%%%%%

\title{
  H$_2$ Emission from the Inner 400 pc of the Galaxy
}
\author{
  Soojong Pak, D. T. Jaffe, and L. D. Keller
}
\affil{
  Department of Astronomy, The University of Texas, Austin, TX 78712
}
\affil{
  E-mail: soojong, dtj, and keller @astro.as.utexas.edu
}

%%%%%%% TODAY'S DATE AND VERSION %%%%%%%
%\author{ ---------------------- }
%\author{ \dateversion }
%%%%%%%%%%%%%%%%%%%%%%%%%%%%%%%%%%%%%%%%

\begin{abstract}

We have mapped the \hh\ \vone\ ($\lambda=2.1215\, \micron$) emission line
along a 400~pc long strip and in a 50~pc region in the Galactic center.
There is \hh\ emission  throughout the surveyed region.
The typical de-reddened ($\ak = 2.5$~mag) \hh\ \vone\ surface brightness, 
$\sim 3 \times 10^{-5}$ \ergintensity, is similar to the
surface brightness in large-scale photon-dominated regions in
the galactic disk. 
We investigate two possible excitation mechanisms for the \hh\ emission:
UV-excitation by photons from OB stars, and shock waves, and conclude
that UV-excitation is more likely.
The total \hh\ \vone\ luminosity in the inner 400 pc region 
of the Galaxy is 8000~\lsun. 
The ratio of the \hh\ to far-IR luminosity in the inner 400 pc of the
Galaxy agrees with 
that in starburst galaxies and ultraluminous infrared bright galaxies.

\subjectheadings{Galaxies:ISM Infrared:ISM:Lines and Bands ISM:Molecules}

\end{abstract}

\section{INTRODUCTION}   \label{sec:int}

Physical conditions in the interstellar medium of 
the Galactic center\footnote{
We use here the term ``Galactic center'' to denote the inner several 100~pc
region of our Galaxy.
We adopt a distance of 8.5~kpc, with which 1\arcdeg\ corresponds to 148~pc.}
are significantly different from those in the solar neighborhood.
The thin disk of interstellar material in the Galactic center 
(size: $450 \times 40$~pc) 
contains $\sim10^{8}\ \msun$ of dense molecular 
gas, $\sim$10~\% of the Galaxy's molecular mass
(G\"{u}sten 1989). 
The molecular clouds in the Galactic center have higher density, 
metallicity, and internal velocity dispersion than 
the clouds in the solar neighborhood (Blitz et al. 1993).
Strong radio continuum radiation from 
giant \hii\ regions 
and extended, low-density (ELD) \hii\ (Sofue 1985),
as well as far-IR radiation from dust (Odenwald \& Fazio 1984),
indicate that the UV radiation field is intense.
The energetic conditions in the Galactic center can
provide a unique view of the interaction between stellar UV radiation and 
molecular clouds, and a nearby example for the nuclei of other 
galaxies.

%Because \hh\ is so light and has no permanent dipole moment,
%its excited rotational states are too high  ($\Delta E / k > 500$~K) to 
%be thermally excited in cold clouds ($T_{kin} < 100$~K).
Ro-vibrational lines of \hh\ trace Photon-dominated Regions (PDRs)
where far-UV photons excite the \hh\ and shocked regions where 
the \hh\ is thermally excited.  As a result,
the central regions ($\sim 1$~kpc) in starburst galaxies are powerful emitters 
of near-IR \hh\ emission
(Puxley, Hawarden, \& Mountain 1989; Joseph 1989; Lester et al. 1990; 
Moorwood \& Oliva 1990).
Vigorous star formation in these galaxies produces large numbers
UV photons which can excite \hh, while subsequent 
supernovae can shock excite the \hh.

Gatley et al. (1984, 1986) and Gatley \& Merrill (1993) 
have observed \hh\ emission
from the inner 5 pc diameter (2\arcmin) in the Galactic Center,
a much smaller region than those observed in starburst galaxies.
With the University of Texas Near-Infrared Fabry-Perot Spectrometer
(Luhman et al. 1995),
it is now possible to observe \hh\ emission over much larger angular scales.
We describe here a program to map the Galactic center in 
\hh\ emission on a scale of several degrees (several hundred pc),
and discuss the likely
\hh\ excitation mechanism.
We can then compare the central region of our Galaxy 
with those in other galaxies.

\section{OBSERVATIONS}   \label{sec:obs}

We observed the \hh\ \vone\ ($\lambda = 2.1215$~\micron)
line at the McDonald Observatory 
0.91~m telescope in 1994 May and June,
using the University of Texas Near-Infrared Fabry-Perot Spectrometer 
(Luhman et al. 1995).
To select a single order from the Fabry-Perot 
interferometer, we used a 1~\% interference filter cooled to 77~K.
The telescope (f-ratio 13.5),
a collimator (effective focal length 343~mm), 
and a field lens (effective focal length 20~mm)
combined to produce a beam diameter of 3\farcm3 (equivalent disk).

The Fabry-Perot interferometer operated in 92nd order
with an effective finesse of 17.7, yielding a spectral resolution 
(FWHM) of 184~\kms.
The scanning spectral range was $\pm 335$~\kms\,
centered at V$_{LSR}$ = 0, with 20 
sequentially exposed channels.
%We aligned the Fabry-Perot etalon 
%every \( 5-8 \) minutes by executing our automatic alignment routine
%(Luhman et al. 1995).
%The Fabry-Perot etalon plates maintained alignment for \( 15-30 \) 
%minutes, but the plate separation drifted by the equivalent
%\( 2-5 \)~\kms\ (wavelength scale) per minute.
%Using the telluric 
%OH $v=9\rightarrow7$ $R_1(1)$ 2.12440~\micron\ line (Oliva and Origlia 1992)
%which was available in each raw spectrum,
%we calibrated the wavelength scale to $\pm 20~\kms$.
%
We nodded the telescope between the object
and the sky every $\sim 60$ seconds to subtract 
background and the telluric OH line emission.
The sky positions were offset by $\pm 1\arcdeg$ in declination 
($\Delta l = \pm 0\fdg85,\ \Delta b = \pm 0\fdg53$)
from the object positions. 
The telescope pointing error was $\pm 15\arcsec$.
We made a strip map along the Galactic plane running across \sgra$^*$, at
$b = -0\fdg05$, from $l = -1\fdg2$ (178~pc) to $+1\fdg6$ (237~pc),
taking spectra at 0\fdg1 or 0\fdg2 intervals (Figure~\ref{fig:gc200pc}).
We also mapped the central 50~pc region including \sgra\ and the radio ``Arc''
(the arched filaments and the vertical filaments,
Yusef-Zadeh, Morris, \& Chance 1984) on a 0\fdg05 grid
(see Figure~\ref{fig:gc50pc}).
The relative flux calibration is accurate to  $\pm$ 15 \~\%.

\section{RESULTS}   \label{sec:res}

%%%%%%%%%%%%% FIGURE 1 %%%%%%%%%%%%%
\begin{figure}[p]
\vbox to 11cm {
  \plotfiddle{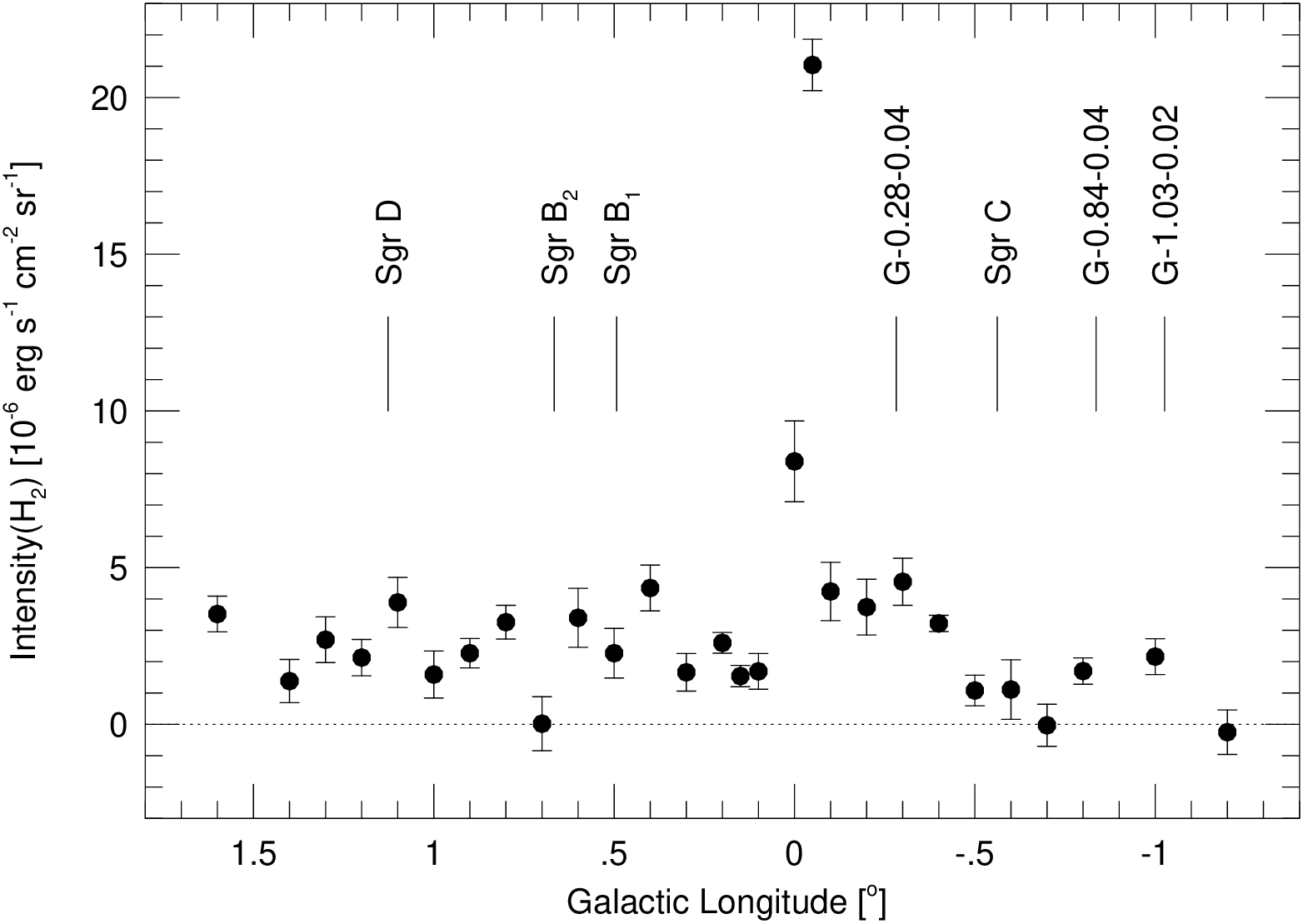} {11cm} {0} {60} {60} {-230} {0}
}
\caption{ \label{fig:gc200pc}
%\figcaption[gc200pc.ps]{ \label{fig:gc200pc}
Observed intensity distribution of \hh\ \vone\ ($\lambda = 2.1215\ \micron$)
along the Galactic Plane at $b = -0^\circ.05$.
The intensity values have not been corrected for interstellar extinction.
The error bars represent $1\sigma$ measurement uncertainties.
The strongest emission at $l=-0\fdg05$ is from \sgra.
The vertical lines give the positions of the prominent radio continuum
sources (Altenhoff et al. 1978).
}
\end{figure}
%%%%%%%%%%%%%%%%%%%%%%%%%%%%%%%%%%%%

%%%%%%%%%%%%% FIGURE 2 %%%%%%%%%%%%%
\begin{figure}[p]
\vbox to 11cm {
  \plotfiddle{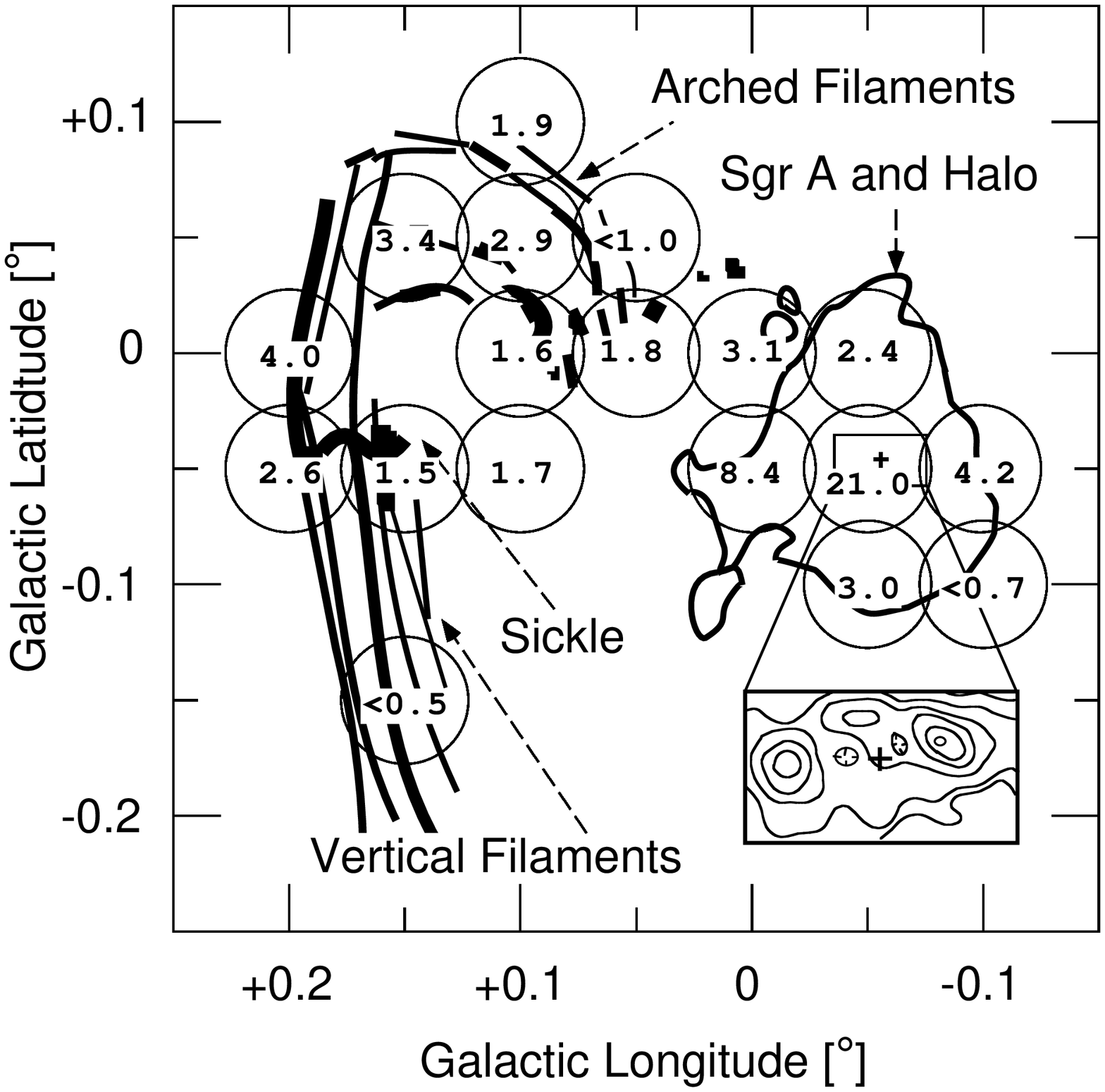} {13cm} {0} {60} {60} {-180} {-20}
}
\caption{ \label{fig:gc50pc}
%\figcaption[gc50pcn.eps]{ \label{fig:gc50pc}
Positions (circles with a diameter of the telescope beam size, 3\farcm3)
and measured \hh\ intensities (numbers inside the circles in units of
10$^{-6}$ \ergintensity)
observed in \hh\ \vone\ in the inner 50~pc of the Galaxy.
Typical measurement uncertainties are 0.7$\times$10$^{-6}$ \ergintensity.
The heavy lines show a schematic version of the radio continuum distribution
(Yusef-Zadeh, Morris, \& Chance 1984).
The small box and inset show the \hh\ \vone\ distribution in
the circum-nuclear gas ring (Gatley et al. 1986).
The plus sign is at Sgr~A$^*$ ($l=-0\fdg0558, b=-0\fdg0462$).
}
\end{figure}
%%%%%%%%%%%%%%%%%%%%%%%%%%%%%%%%%%%%
  
There is \hh\ emission throughout the 400~pc 
diameter region around the Galactic Center.
About 70~\% of the observed positions along the strip at $b=-0\fdg05$ 
have detections of the \hh\ \vone\ 
line with a significance of $3\sigma$ or more.
Figure~\ref{fig:gc200pc} 
shows the intensity distribution of \hh\ \vone\
along the strip.
The \hh\ intensity peaks strongly at \sgra\ and decreases continuously
from $l = -0\fdg1$ to $-0\fdg7$.
The ``dust ridge'' seen in 800~$\mu$m continuum 
emission (Lis \& Carlstrom 1994) may cause 
the dip between $l = +0\fdg1$ to $+0\fdg3$.
Away from the nucleus, the intensity distribution is fairly flat.
%There is no positive longitude enhancement in \hh\, while observations of
%[\ci] $^3$P$_1 \rightarrow ^3$P$_0$, $^{13}$CO $J=1 \rightarrow 0$ (Plume et al. 1995),
%and radio continuum (Mezger \& Pauls 1979) show significantly more emission at 
%the positive longitudes.

Figure~\ref{fig:gc50pc} shows the \hh\ \vone\ intensity distribution 
in the inner 50~pc of the 
Galaxy overlayed on a simplified radio continuum map
(based on Yusef-Zadeh, Morris, \& Chance 1984).
We have detected \hh\ emission in the arched filaments and in the 
Sickle at levels comparable to those of the \hh\ 
emission elsewhere along the Plane.
The inset in Figure~\ref{fig:gc50pc} shows the \vone\ distribution in the 
$0\fdg022 \times 0\fdg038$ region around Sgr~A~West mapped by
Gatley et al. (1986).
%observed strong near-IR \hh\ emission lines, i.e.,
%\hh\ \vone, \vtwo, and \hbox{$v = 1 \rightarrow 0$ S(0)},
%around the Galactic nucleus ($l=-0\fdg0558, b=-0\fdg0462$).
In Gatley's work,
the \hh\ appears to be brightest along the inner edge of 
the circum-nuclear gas ring at radius of $1.0 - 2.5$~pc.
%(see the inset in Figure~\ref{fig:gc50pc}).
Our measured flux at ($l=-0\fdg05$, $b=-0\fdg05$),
$1.5 \times 10^{-11}$ \ergflux, agrees within the errors with
the total flux from the map of Gatley et al. (1986,
$F = 2.0 \times 10^{-11}$ \ergflux).
The \hh\ emission observed adjacent to
($l = -0\fdg05$, $b = -0\fdg05$)
most likely arises from portions of the circum-nuclear ring beyond their map. 

\section{ DISCUSSION } \label{sec:dis}

\subsection{ Extinction Correction } \label{sec:extinction}

At 2.2 $\mu$m, the emission from the Galactic center is attenuated by
interstellar material in the foreground (``foreground extinction'',
mostly 4-8 kpc from the Galactic Center) and by material in the 
Galactic center itself (``Galactic center extinction'').  
Catchpole et al. (1990) mapped the  extinction toward the Galactic
center by observing the near-IR reddening of giant stars in the
central few hundred parsecs.  Along our H$_2$ strip at b= --0\fdg05
(for --0\fdg6$<l<$+0\fdg6), the extinction is fairly uniform with a value 
A$_K\sim$2.5 mag.  Although most of this extinction is in the foreground,
some of it could occur within the Galactic center since Catchpole
et al. (1990) were able to identify patches in their maps with
A$_K>$2.5 with known molecular clouds in the Galactic center
(see plate 4 in Glass et al. 1987).  Based on
this work, we adopt A$_K$=2.5 for the foreground extinction.

The Galactic center extinction greatly exceeds the foreground
extinction.  Typical $^{12}$CO J=1$\rightarrow$0 linestrengths
along the strip we have surveyed in H$_2$ are 1500 Kkms$^{-1}$
(Oort 1977).  This line strength implies an A$_K$ of 10--40 mag,
depending on the CO/H$_2$ and A$_K$/H$_2$ ratios in the Galactic
center (Sodroski et al. 1994).  The extinction through individual clouds
may also be substantial (A$_K\sim$10--30 for a 10 pc long cloud with 
n$_{H_2}$=10$^4$ cm$^{-3}$). 
The relevance of the Galactic center extinction depends on the source
of the H$_2$ emission.  Any H$_2$ emission originating within the clouds 
will be highly extincted.  If the H$_2$ emission arises on the
cloud surfaces, however, we only miss the H$_2$ flux from the back
side of each cloud.  

Clouds lying in front of other clouds will further reduce the flux
reaching us from the front surfaces.
If, in the Galactic Center, the velocity integrated area 
filling factor of clouds, {\it f},
is substantially greater than unity, 
extinction by overlapping clouds will reduce the H$_2$ flux observed
from the front surfaces
by a factor $\sim1/f$ in addition to the foreground extinction
and to the loss of emission on the opposite sides of the clouds.
Typical clouds in the Galactic center disk have kinetic temperatures
$\sim$70 K and linewidths $\sim$20 km s$^{-1}$ (G\"usten 1989).
An ensemble of such clouds could produce the observed 
$^{12}$CO J=1$\rightarrow$0 lines in the Galactic center with f$\sim$1.
We therefore conclude that the extinction of any H$_2$ emission from
cloud surfaces facing the sun beyond the foreground extinction of
A$_K$= 2.5 discussed above
is not substantial.  Since the extinction of emission from within the
clouds or from the sides facing away from us is difficult to estimate
and since no correction is usally made for such effects in 
giant molecular clouds and galactic nuclei, we make no additional
extinction corrections here.

\subsection{ UV Excitation of  \hh\ } \label{sec:UV}

If one ignores the region immediately around \sgra$^*$,
the de-reddened ($\ak = 2.5$~mag, see Section~\ref{sec:extinction}) 
\hh\ \vone\ surface brightness along the Galactic 
plane has a roughly constant value of
$\simeq 3 \times 10^{-5}$ \ergintensity.
Any excitation mechanism for the \hh\ must be able to explain both the 
absolute intensity and the uniformity and extent of the emission.
The excitation of the \hh\ $v=1$, $J=3$ state can result either from radiative
decay from UV-excited electronic states or from energetic collisions.
\hh\ can absorb 91--123 nm photons in the $B^1\Sigma^+_u - X^1\Sigma^+_g$
Lyman and $C^1\Pi_u - X^1\Sigma^+_g$ Werner bands.  About 90\%
of the time, the excited \hh\ decays to some ro-vibrational level of the ground 
electronic state.  The relative line intensities arising in UV-excited \hh\
are insensitive to density or to UV field strength if 
${\rm n_{H_2}} < 10^4$ \cmv\ (Black and van Dishoeck 1987).  
At densities $\geq 10^5$ \cmv,
UV-excited gas can become hot enough that collisions populate states with
$v=1$ (Sternberg and Dalgarno 1989).  Collisional excitation can also result 
from shocks which abruptly heat the gas to $> 10^3$~K (e.g. 
Hollenbach, Chernoff, \& McKee 1989).  
Several observational results lead us to believe that
UV excitation can explain the \vone\ emission in the Galactic Center.

The denser parts of clouds like Orion and NGC 2024 
produce \hh\ emission with an intensity close to that observed 
in the Galactic Center.
In Orion and
NGC 2024, the degree-scale H$_2$ emission has a typical surface brightness
$\sim 6 \times 10^{-6}$ \ergintensity\ (Luhman et al.
1994).  Along the molecular ridges in these clouds, the H$_2$ 
surface brightness is 3-5 times higher. 
Observations of \hh\ transitions
arising from higher-lying states indicates that,
in these sources, the \vone\ emission
is a result of UV fluorescence.  
%In Orion, the UV-excited emission accounts for at least
%90\% of the total flux in the \vone\ line (Luhman et al. 1994).    

If large-scale \hh\ emission arises in the surface layers of the clouds
where UV photons can excite the molecules, the dust, which absorbs the
bulk of the incident flux, ought to radiate in the far-IR continuum as well.
Luhman \& Jaffe (1996) have compared the \hh\ \vone\ observations of clouds
in the galactic disk with IRAS far-IR continuum results and derived
a relation between the \hh\ \vone\ line and far-IR continuum intensities.
We can use this relationship and the measured far-IR intensities
in the Galactic center to see if UV-excitation is plausible for our 
observed \vone\ emission.
In most of the region along our Galactic center H$_2$ cut, 
the IRAS 100 $\mu$m band detectors were saturated.  In order to compare the
Galactic center H$_2$ data to far-IR continuum measurements with comparable
angular resolution, we have combined the un-saturated IRAS measurements
from the outer ends of our H$_2$ strip with the 40--250 $\mu$m continuum
measurements of Odenwald and Fazio (1984). To make the two data sets 
comparable, we have first converted the IRAS 60 $\mu$m and 100 $\mu$m
fluxes into a total far-IR
flux (the FIR parameter of Fullmer and Lonsdale 1989).  The IRAS total
far-IR flux agrees well with the far-IR flux derived by Odenwald and
Fazio in the regions where their data and the unsaturated IRAS data overlap.
We then converted the combined datasets into integrated far-IR
intensity for comparison
with our H$_2$ strip.
We used the Luhman \& Jaffe galactic disk \hh\
dataset to re-derive their H$_2$/far-IR
relation in intensity units.  We obtain,
\begin{displaymath}
  \log(I_{\rm H_2 v=1-0 S(1)}) = -4.65 + 0.39 \log(I_{FIR}),
\end{displaymath}
where both intensities are in \ergintensity.  
The dispersion of the galactic disk cloud \hh\ intensities about this relation
is log($\sigma$) = 0.23.  If we de-redden the Galactic center H$_2$ observations
by \ak\ = 2.5~mag ({\it but otherwise do nothing to fit the data to 
the galactic disk relation}), the Galactic center \hh\
intensities have a dispersion log($\sigma$) = 0.26 about this relation. 
The Galactic center results are therefore completely consistent
with the empirical far-IR vs. \hh\ relationship derived for the 
UV-excited surfaces of clouds in the galactic disk. 
 
We can also compare the H$_2$ line intensities predicted by
models of photon-dominated regions to the observed intensities.
The models use indirect observations of the far-UV field in the
Galactic center (radio and far-IR continuum fluxes) as inputs.
For the radio, we predict the far-UV field using emission from
extended, low-density (ELD) \hii\ regions because the molecular cloud 
column densities, and therefore the extinction at the wavelength of \hh,
tend to be high (and uncertain) 
toward the discrete \hii\ regions.
Away from discrete H~II regions, the typical 10.5 GHz flux density is 2.2 Jy
in a 3\farcm3 beam (Sofue 1985). 
Assuming $T_e = 10^4$~K, this flux density corresponds to 
$2.3 \times 10^{49}$~sec$^{-1}$ Lyman continuum photons per second
(Mezger, Smith, \& Churchwell 1974), 
in the corresponding region (8.2~pc). 
For an ionizing source with an effective stellar temperature, 
$T_{eff} = 3.5 \times 10^4$~K as the UV source, 
the 2.3$\times$10$^{49}$ Lyman continuum
photons imply $\sim$2.3$\times$10$^{49}$ photons in the range 
which can excite the \hh\ (91-123 nm), or a luminosity of  
$1.2 \times 10^5$~\lsun\ (Black \& van Dishoeck 1987).
From our observations, the average \hh\ flux in a 3\farcm3 beam is
$2.4 \times 10^{-12}$~\ergflux.
The corresponding total \hh\ luminosity in the 8.2~pc (3\farcm3) region 
is $3.3 \times 10^3$~\lsun,
if we correct for an extinction of $\ak = 2.5$~mag and use the PDR model of
Black \& van Dishoeck (1987) to extrapolate to the \hh\ cooling in all 
lines ($I_{\rm H_2 v=1-0 S(1)}/I_{\rm H_2}$ $=$ $0.016$).
The ratio of the near-IR \hh\ luminosity to the luminosity in the
far-UV band 
that is effective in exciting \hh\ is 0.028,
which is close to the value of 0.034 from an appropriate PDR model for the
Galactic center (Model 19 in Black \& van Dishoeck 1987, which has
n$_H$ = 10$^4$ and a UV field I$_{UV}$ = 10$^3$). 
%Therefore the UV-excitation can explain the observed \hh\ strength.

The far-IR continuum intensities along our Galactic center strip are
typically 0.8 \ergintensity\ (Odenwald and Fazio 1984).  If all of this
emission arises from a single molecular cloud surface filling the beam,
it corresponds to a far-UV flux $\sim 2 \times 10^3$ times the mean
interstellar radiation field in the solar neighborhood (Draine 1978).
Given the likely number of clouds along each line of sight and various
geometric effects, the likely far-UV field is $500-1000$ times the solar
neighborhood value.  For this range of UV field strengths and densities
between 3$\times$10$^3$ and 3$\times$10$^4$ \cmv,  Black and 
van Dishoeck (1987) predict \hh\ \vone\ line intensities in the range
$1.2-4.2 \times 10^{-5}$ \ergintensity, bracketing our typical observed,
de-reddened value.

The \hh\ emission from the circum-nuclear disk appears to be collisionally 
excited 
(I$_{\rm v=2-1 S(1)}$/I$_{\rm v=1-0 S(1)}$) $\simeq 0.1$, Gatley et al. 1984).
Gatley et al. suggest that shocks created by mass outflow 
from the Galactic nucleus might excite the \hh.
Such thermal line ratios can also occur, however, in UV-excited gas if the
UV fields and densities are sufficiently high (Sternberg and Dalgarno 1989;
Luhman et al. 1996).
%The line ratio, however, does not prove that the molecular gas is 
%heated dynamically by shock waves.
%Quiescent dense neutral gas, which is heated radiatively by 
%UV photons, emits thermal \hh\ line radiation.
Since the typical hydrogen density in the 
circum-nuclear disk is large, i.e.,
$n_H \simeq 10^5\ \cmv$, and
the UV field is intense in the central 3~pc, 
(the number of total Lyman continuum photons absorbed by the gas is
$\sim2 \times 10^{50}\ \lsun$,  Lacy et al. 1980), the strength and
character of the \hh\ emission from the circumnuclear disk are
also consistent with UV-excitation.

\subsection{ Shock-Excitation }   \label{sec:shock}

Shock excitation of the \hh\ \vone\ transition must take place, at some level,
in the inner 400 pc of the Galaxy.
A large variety of dynamical activity may give rise to shocks with
appropriate characteristics.  Outflows around newly formed stars and
shocks caused by supernova remnants impinging on molecular clouds
in the galactic disk both produce \hh\ emission and should be observable
in the Galactic Center. 
Bally et al (1987; 1988) surveyed the Galactic center region in 
the $^{12}$CO and \thirteenco\ $J=1 \rightarrow 0$, and 
CS $J=2 \rightarrow 1$ lines.
The gas distribution is highly asymmetric about the center, and
some negative velocity gas is seen at positive 
longitudes, which is ``forbidden'' to gas in circular orbits.
This gas and other clouds with eccentric orbits may
collide with material in more circular orbits.
For example, in the \sgrbtwo\ complex, 
Hasegawa et al. (1994) suggested that
 a dense ($ n_{H_2} \simeq 1.4 \times 10^4\ \cmv $), 
 $ 10^6\ \msun $  ``Clump'' 
%($l=+0\fdg68$, $b=-0\fdg03$)
 has 
collided with the extended less dense ``Shell'' of the cloud complex
producing a $\sim 30\ \kms$ shock.  
Finally,
the internal velocity dispersion of the molecular clouds is in the range of
$\Delta V = 20 - 50\ \kms$ (Bally et al. 1988).
If the internal collisions efficiently 
dissipate the relative kinetic energy by radiative cooling following shocks, 
there should be \hh\ emission throughout the 
molecular clouds, much of it, however, heavily extincted.

Depending on the context, shock-excited \hh\ emission could result either
from dissociative J-shocks (colliding clouds, supernova remnants), or from
C-shocks (outflows, dissipation of turbulence).  
The J-shocks give rise to \hh\ \vone\ intensities in the range
of 3$\times$10$^{-5}$ -- 10$^{-4}$ \ergintensity\ with the intensity being
fairly insensitive to density and shock velocity over the range
10$^4$ cm$^{-3} \leq$ n $\leq$ 10$^5$ cm$^{-3}$ and 
30 km s$^{-1} \leq$ v$_{shock} \leq$ 150 km s$^{-1}$ 
(Hollenbach \& McKee 1989). For A$_K$ = 2.5~mag, the predicted intensity
matches what we observe in the Galactic center fairly well.
In order to explain the distribution of observed \hh\ emission, however,
the number of shock fronts times the area covered per beam must roughly
equal the beam area along virtually every line of sight through the inner
400 pc of the Galaxy, an unlikely picture at best. 

C-shocks can produce \hh\ \vone\ intensities in 
the range of those observe in the Galactic Center.  A single C-shock with
n = 10$^4$ cm$^{-3}$ and V = 20 km s$^{-1}$ gives
I$_{S(1)}$ $\simeq$ 3$\times$10$^{-5}$ \ergintensity\ (Draine, Roberge, \&
Dalgarno 1983).  The emergent intensity, however, is extremely sensitive
to the shock velocity, varying (at n$_H$ = 10$^4$ cm$^{-3}$)
by 3 orders of magnitude from V$_{shock}$ = 18 to V$_{shock}$ = 40~\kms .
A model making use of C-shocks to produce the observed uniform \hh\
\vone\ distribution would have to be somewhat contrived.
While there may be some shock-excited
\hh\ emission from the Galactic Center, 
it is difficult to argue away the expected PDR emission
and then construct a reasonably simple shock model capable of explaining 
the observations.
A reliable test of the excitation mechanism would be to observe \hh\
transitions arising higher above ground than the \vone\ line.

\subsection{ Total \hh\ Luminosity }

To estimate the total \hh\ luminosity of the Galactic Center, 
we extrapolate from our 400 pc long strip by
assuming that the scale height 
of the \hh\ emission equals that of the far-IR radiation
($h\simeq 0\fdg2$, Odenwald \& Fazio 1984).
For $\ak = 2.5$~mag and $f \leq 1$ (see Section~\ref{sec:extinction}), 
the de-reddened \hh\ \vone\ luminosity in the inner 400~pc diameter 
of the Galaxy is $8.0 \times 10^3$~\lsun. 
Joseph (1989) gives ranges of \hh\ \vone\ luminosity in $> 1$~kpc regions 
for various classes of galaxies: 
(1) merging galaxies: $3 \times 10^6 - 3 \times 10^8\ \lsun$;
(2) interacting galaxies: $10^5 - 10^7\ \lsun$;
(3) barred spirals: $10^4 - 10^6\ \lsun$.
Over its inner $\sim 1$~kpc,
our Galaxy most likely falls within the range for barred spirals.

In ultraluminous 
infrared bright galaxies ($L_{IR} \gtrsim 10^{12}\ \lsun$),
Goldader et al. (1995) show that 
log($L_{S(1)}/L_{FIR}$) $=$ $-4.95 \pm 0.22$.
For the nearby starburst M82. we can use \hh\ \vone\ measurements of 
the inner 60\arcsec\ (Pak \& Jaffe, unpublished) together with far-IR
continuum observations (D. A. Harper, as quoted in Lugten et al. 1986) to derive
log(L$_{S(1)}$/L$_{FIR}$) = $-5.2$ for the inner 1 kpc.
For the inner 400 pc of the Milky
Way, the data presented here yield log(L$_{S(1)}$/L$_{FIR}$) = $-4.8$.

There is evidence in some high-luminosity galaxies that the \hh\ 
emission results from UV-excitation.
In NGC~3256, a merging galaxy,
the \hh\ \vtwo /\vone\ line ratio 
in the 600~pc region ($3\farcs5 \times 3\farcs5$) is $0.39 \pm 0.06$, 
suggesting that UV fluorescence is responsible for at least 60~\% of
the \hh\ \vone\ emission (Doyon, Wright, \& Joseph 1994) .
If \hh\ in the Galactic center is UV-excited, as we suggest here, this 
mechanism could be shared by \hh\ emission from galaxies with an enormous
range of nuclear conditions.

%\section{ CONCLUSIONS }
%
%We have surveyed the inner 400~pc diameter of the Galaxy 
%in the \hh\ \vone\ emission line. 
%The emission is extremely widespread and has a typical dereddened intensity
%of 3$\times$10$^{-5}$ \ergintensity. Empirical and theoretical models
%of PDR's both indicate that UV-excitation can explain the strength and 
%distribution of the \hh\ emission.  Observations of extended \vsix\ emission in %Orion unambiguously
%demonstrate that the extended \hh\ in that source is UV-excited
%(Luhman et al. 1994).  Comparable observations in the Galactic center 
%would be difficult but are possible.
%
%
%The total \hh\ \vone\ luminosity in the 400~pc of the Galaxy is 
%$8.0 \times 10^3\ \lsun$,
%an order of magnitude larger than that in the strong and compact 
%circum-nuclear gas ring ($7 \times 10^2\ \lsun$).
%The luminosity ratio of the \hh\ \vone\ to far-IR in the Galaxy agrees 
%with that in starburst galaxies and in ultraluminous infrared bright
%galaxies.

\acknowledgments

This work was supported by the David and Lucile Packard Foundation
and by NSF grant AST 9117373.  We thank 
John Lacy and the referee, Leo Blitz for helpful comments, 
Michael Luhman and Al Mitchell for 
contributions to the Fabry-Perot Spectrometer Project, and Jacqueline Davidson 
and the staff of the McDonald Observatory for their assistance on the 
observing run.

%%%%%%%%%%%%%%%%%%%
%\clearpage

%%%%%%%%%%%%%%%%%%%%%%%%% References %%%%%%%%%%%%%%%%%%%%%%%%%%%%

\end{document}